\documentstyle[multicol,prl,epsf,aps]{revtex}   

\begin{document}

\title{A Simple model for Faraday Waves}

\author{John Bechhoefer$\footnote{e-mail: bech@chaos.phys.sfu.ca}$}

\address{Department of Physics, Simon Fraser University, Burnaby, B.C., 
V5A 1S6 Canada}

\author{Brad Johnson}

\address{Department of Physics, Western Washington University, Bellingham, 
WA 98225}

\maketitle

\begin{abstract}
We show that the linear-stability analysis of the birth of Faraday waves on
the surface of a fluid is simplified considerably when the fluid container
is driven by a triangle waveform rather than by a sine wave.  The
calculation is simple enough to use in an undergraduate course on fluid
dynamics or nonlinear dynamics. It is also an attractive starting point for 
a nonlinear analysis.
\end{abstract}

\pacs{47.20.-k, 47.35.+i}

\begin{multicols}{2}
\narrowtext

\section{Introduction}
\label{sec:intro}

	Parametrically excited surface waves, or Faraday waves, are
generated when a container of fluid is vertically vibrated.  Above an
acceleration threshold, waves appear on the fluid surface.  The waves
vibrate at half the driving frequency, a feature associated with the
parametric resonance that is responsible for pumping energy into the
surface-wave modes.

	In most ``Faraday experiments,'' the container is driven
sinusoidally.  That is, its vertical position may be described by $x(t) =
(x_{pp}/2) \cos\omega t$, and its acceleration by $a(t) = - \omega^2
(x_{pp}/2) \cos\omega t$.  In this paper, we explore the consequences
of driving the container with a triangle-wave forcing.  Although the
difference between the two waveforms matters little physically, we shall
show that it simplifies the analysis remarkably.

	There are two broad motivations for our study.  First, although the
theory of hydrodynamic stability occupies an important conceptual place in
current courses in hydrodynamics, standard examples such as thermal
convection (Rayleigh-B\'enard Convection, or RBC) and shear flow between
rotating concentric cylinders (Taylor-Vortex Flow, or TVF) are too
complicated to present in standard undergraduate courses.  Textbooks for
such courses (e.g., \cite{tritton88}) usually limit themselves to
heuristic arguments.  By contrast, our analysis is simple enough to present
in a single class to third- or fourth-year students.  (The calculation's
difficulty is equivalent to solving the Kronig-Penney model for energy bands
in a one-dimensional solid.  This example is commonly presented in
undergraduate solid-state-physics courses.)

	A second motivation comes from the increasingly prominent role that
the Faraday experiment occupies as an example of a non-equilibrium
pattern-forming system.  Most of the work on pattern formation done to date
has focused on convection (RBC) and shear flow (TVF) \cite{cross93}.
However, a recent series of papers has established a number of advantages
for the Faraday experiment 
\cite{ezersky86,christiansen92,fauve92,edwards94,bosch93,gluckman95,daudet}.  
Very large cells ($> 10^2$ wavelengths wide) may be created, and the 
basic (``fast") time scale can
be very short ($t_0 \lesssim 10^{-2}$ sec.)  The short time
scale is particularly important because the dynamics of pattern-forming
instabilities is often slow.  Current work looks at features that occur on
time scales of $10^4 t_0$ to $10^5 t_0$.  Convection, by contrast, usually
has $t_0 \approx 1$ sec., or slower.

	As we shall show below, our variant of the Faraday experiment
clarifies our understanding of the instability itself.  The analysis
separates into two pieces:  one is the derivation of the surface-wave
dispersion relation (Section III); the other is the description of 
how parametric pumping injects energy into the fluid (Section II).  
By choosing triangle-wave forcing, we
greatly simplify the latter piece.  We hope that this will ultimately allow
the nonlinear analysis (particularly of high-viscosity fluids) to be pursued
further than it has been hitherto.  For this reason, we make no restriction
on the fluid viscosity; the only approximation made is in linearizing the
Navier-Stokes equations.

	The body of this paper is organized as follows:  In Section
\ref{sec:mathieu}, we study the related problem of the Mathieu equation
with delta-function forcing.  Solving this problem turns out to be
mathematically equivalent to solving for the acceleration thresholds of
Faraday waves.  In Section \ref{sec:faraday}, we apply this analysis to the
Faraday experiment with triangle-wave forcing.  In Section \ref{sec:2freq},
we consider a simple extension, driving by an asymmetric triangle wave, 
which models the recent ``two-frequency" experiments that have attracted 
much attention \cite{edwards94}.  Section \ref{sec:conclusion} contains 
a brief conclusion.

\section{The Mathieu Equation with Delta-Function Forcing}
\label{sec:mathieu}

	An infinitely extended continuous system such as a fluid can be 
thought of as containing a continuum number of modes.  To linear order, the
modes are independent and the system's behavior can be determined from the
study of each mode separately.  Nonlinear effects imply mode coupling.  In
this section, we examine the effects of parametric forcing on a single mode.
In Section \ref{sec:faraday}, we apply the results derived here to the
continuum of modes present in the Faraday experiment.

	There are two ways to pump energy into an oscillator.  The first is
by direct forcing, where the prototypical equation is 
\begin{equation}
\ddot{x} + 2 \gamma \dot{x} + \omega_0^2 x = f(t) ,
\end{equation}
where $f(t)$ is the driving force and is often chosen to be $f \cos\omega t$.  
For small damping $\gamma$, a directly forced oscillator has a resonant 
response (of amplitude $\approx f/\omega_0\gamma$) at 
$\omega \approx \omega_0$.

	The second way to pump energy into an oscillator is by parametric
forcing, whose prototypical equation is the (damped) Mathieu equation:
\begin{equation}
\ddot{x} + 2 \gamma \dot{x} + \omega_0^2 [1+f(t)]x = 0 .
\label{eq:mathieu}
\end{equation}
When $f(t)$ is sinuosoidal, there is an infinite series of resonances
centered on $\omega = 2\omega_0$, $\omega_0$, $(2/3)\omega_0$, 
$\omega/2$, $... , (2/n) \omega_0$, for integer $n$.  
The strongest of these resonances is the first
($n=1$), and a response at half the driving frequency is good evidence that
a system is being driven parametrically.  Below, we shall find it convenient
to rephrase the resonance conditions in terms of periods:  $\Delta t = (n/2)
T$, where $\omega = 2\pi /\Delta t$ and $\omega_0 = 2\pi /T$.

\begin{figure}[tbp]
\epsfxsize=6cm
\epsffile{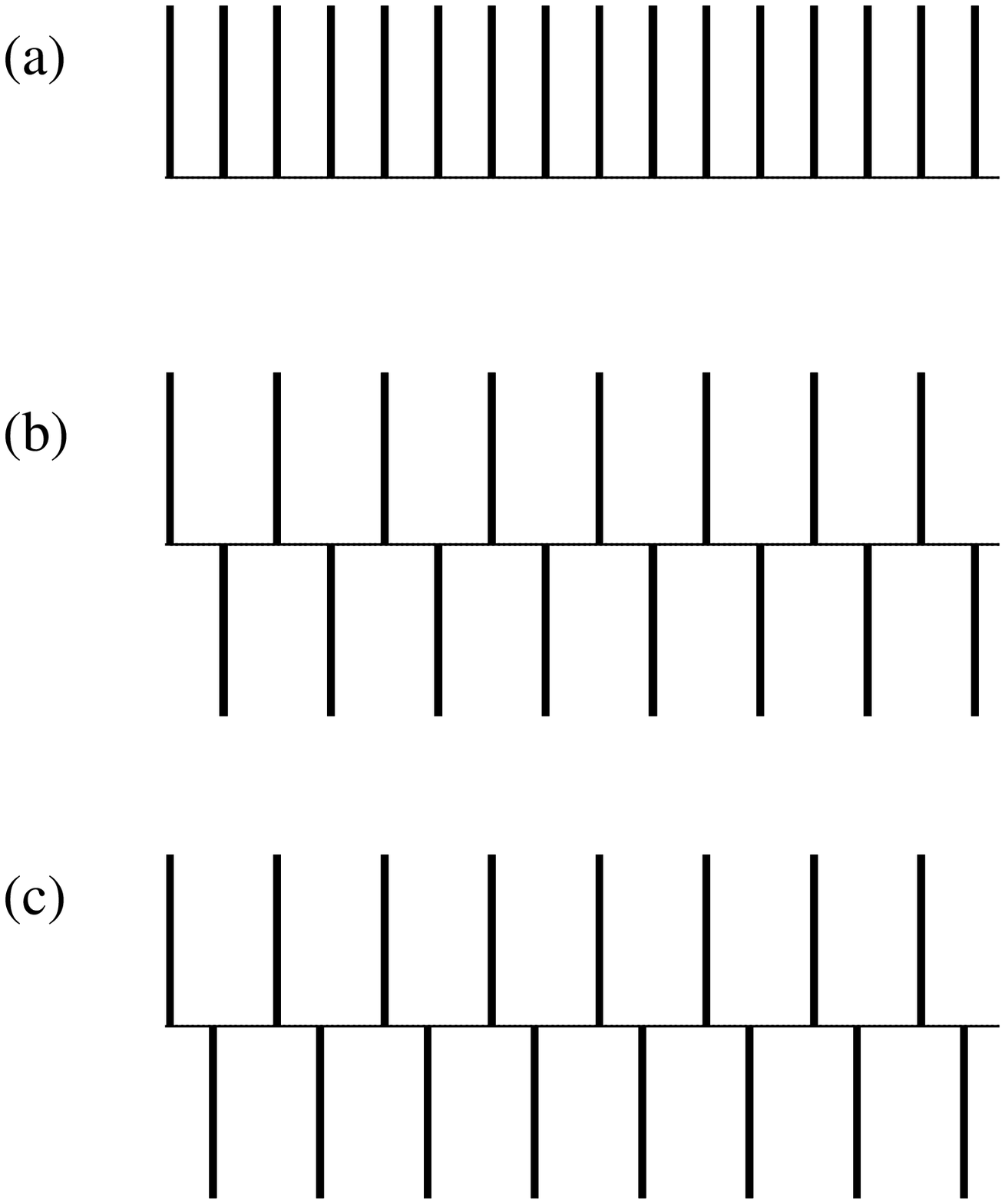}
\caption {Driving function $f(t)$. (a)  Periodic train of delta functions of
equal signs.  (b)  Periodic train of delta functions of alternating signs,
derived from a symmetric triangle wave. (c)  Periodic train of delta 
functions of alternating sign, derived from an asymmetric triangle wave.}
\label{fig:driving}
\end{figure}

	Although $f(t)$ may be sinusoidal, the solutions $x(t)$ are not.  
Let $f(t) = f \cos 2\omega_0 t$ and
assume $x(t) = x_0 \cos\omega_0 t.$  The term $f(t)x(t)$ in 
Eq. (\ref{eq:mathieu}) is proportional to 
$\cos\omega_0 t \cos 2\omega_0 t$ and can be rewritten as 
$(1/2)(\cos \omega_0 t + \cos 3\omega_0 t)$, 
implying that $x(t)$ has a $3\omega_0
t$ component.  Adding this to $x(t)$ generates a $5\omega_0 t$ term, 
and so on.  Eq. (\ref{eq:mathieu}) must in general be solved numerically.
For special choices of $f(t)$, however, Eq. (\ref{eq:mathieu}) may be 
solved analytically.  The simplest case
is $f(t)$ equal to a sum of delta functions \cite{hsu72}.  A slightly
more complicated case, $f(t)$ equal to a square wave, was discussed in these
pages some years ago \cite{yorke78}.  In Section \ref{sec:faraday}, we show
that analyzing a fluid container driven by a triangle wave is equivalent to
studying the Mathieu equation driven by a series of equal-amplitude delta
functions, with alternating signs.  (See Fig. \ref{fig:driving}b.)  Here, we
review the analysis of \cite{hsu72} for the particular case of delta-function
forcing with equal and then with alternating signs.

\subsection{Mathieu equation driven by a periodic series of delta functions}
\label{sec:equalsigns}
 
In units where time, damping rate, and forcing are all scaled by $\omega_0$,
we wish to solve
\begin{equation}
\ddot{x} + 2 \gamma \dot{x} + [1+\varepsilon \sum_n \delta (t-n\Delta t)]x = 0 .
\label{eq:delta1}
\end{equation}
The forcing function is shown in Fig. \ref{fig:driving}a.
In between delta-function ``kicks," the system is just a free, damped
oscillator.  Let $x_n (t)$ be the solution valid between the kicks at
$n\Delta t$ and $(n+1)\Delta t$.  One finds
\begin{equation}
x_n(t) = A_n e^{\psi (t-t_n)} + \rm{c.c.}
\label{eq:solution}
\end{equation}
with $\psi = - \gamma \pm i \omega_0'$ and $\omega_0' = \sqrt{1 -
\gamma^2}$.  The kick imposes two conditions linking $x_n$ and $x_{n+1}$:

	$\bullet$ $x$ is continuous. 

	$\bullet$ the velocity $\dot{x}$ jumps discontinuously.  This condition
can be derived by integrating Eq. (\ref{eq:delta1}) over a small 
interval of time centered on $t_{n+1}$.  

	The conditions may be expressed as:
\begin{mathletters}
\label{eq:conditions}
\begin{eqnarray}
x_{n+1} (t_{n+1}) &=& x_n (t_{n+1}) \\
\dot{x}_{n+1} (t_{n+1}) &-&
\dot{x}_n (t_{n+1}) + \varepsilon x_n (t_{n+1}) = 0 .
\end{eqnarray}
\end{mathletters}
Imposing the continuity and velocity-jump conditions allows 
one to relate $A_{n+1}$ to $A_n$:
\begin{equation}
\left(
\begin{array}{c}
A_{n+1}^r \\
A_{n+1}^i
\end{array}
\right) = e^{-\gamma \Delta t} \left(
\begin{array}{cc}
C & - S \\
S + \alpha C & C - \alpha S
\end{array} \right)
\left(
\begin{array}{c}
A_n^r \\
A_n^i
\end{array}
\right) , 
\label{eq:An}
\end{equation}
where $A_n^r$ and $A_n^i$ are the real and imaginary parts of $A_n$,
respectively, $C = \cos\omega_0'\Delta t$, $S = \sin\omega_0'\Delta t$, and
$\alpha = \varepsilon /\omega_0'$ is the scaled forcing amplitude. 

	Eqs. (\ref{eq:solution}) and (\ref{eq:An}) constitute a solution to 
Eq. (\ref{eq:delta1}).  Given initial conditions $A_0$, we can find $x(t)$ 
by first iterating the map $e^{-\gamma\Delta t} M$ to find
the proper $A_n$.  Here, $M$ is the $2 \times 2$ matrix in Eq. (\ref{eq:An}). 
Then use Eq. (\ref{eq:solution}) to find $x(t)$.

	Eq. (\ref{eq:An}) can also be used to find the acceleration threshold
for resonant response.  Parametric resonance differs from
ordinary resonance in that it is a true instability.  If $\varepsilon$ is
below a threshold $\varepsilon_c$, $x(t) \to 0$ at long times.  For
$\varepsilon > \varepsilon_c$, $x(t)$ grows without bounds.  (Adding a
nonlinear term in $x$ to Eq. (\ref{eq:delta1}) will limit the oscillator's
amplitude.  In practice, nonlinearities are always present in the physical
situation being modeled.)  Our goal now is to find 
$\varepsilon_c({\Delta t})$.

	The key observation is that the long-time amplitude of the motion
$x(t)$ is controlled by the eigenvalues of the matrix $M$: if their
magnitude is greater than $e^{\gamma\Delta t}$, the motion's
amplitude will grow each period.  If the magnitude of both eigenvalues is less
than $e^{\gamma\Delta t}$, damping will reduce the motion each cycle and the
system will tend towards $x(t) = 0$.  Loosely speaking, $M$ accounts for the
energy pumped into the system during each period $\Delta t$, while
$e^{-\gamma\Delta t}$ represents the energy lost during $\Delta t$.

\begin{figure}[tbp]
\epsfxsize = 6cm
\epsffile{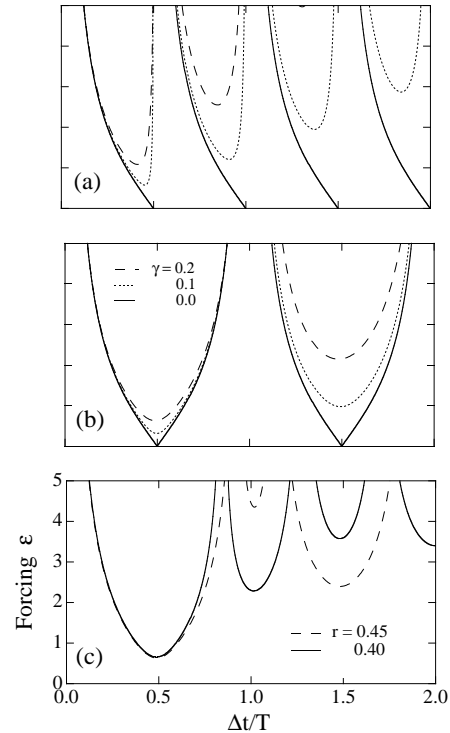}
\caption {Resonance tongues for the Mathieu eq. driven by a periodic sequence
of delta functions, for $\gamma$ = 0.0, 0.1, and 0.2. (a) Delta functions of equal sign.  (b) Delta functions of
alternating signs.  (c)  Delta functions derived from an asymmetric triangle 
wave, for $\gamma = 0.2$ and different values of $r \equiv \Delta t_2/
\Delta t$.  For $r = 0.5$, the tongues are the same as the dashed lines in
(b).  Note how the harmonic tongues are more prominent as the asymmetry is
increased ($r \to 0$).}  
\label{fig:delta_tongues}
\end{figure}

	The matrix $M$ has unit determinant.  If we write $M$ in the general
form 
$\left( \begin{array}{cc}
A & B \\
C & D
\end{array} \right)$
the eigenvalues are given by
$\lambda_{1,2} = x \pm \sqrt{x^2 -1}$
where $x = (A+D)/2 = (1/2)$Tr$M$.  Instability arises when both
eigenvalues are real and the magnitude of the largest exceeds $e^{\gamma\Delta
t}$.  Simple algebraic manipulations then give for the threshold condition
\begin{equation}
x = \pm \cosh \gamma\Delta t
\label{eq:threshold}
\end{equation}
In the present case, the matrix elements are $A = \cos\omega_0' \Delta t$
and $D = \cos\omega_0' \Delta t - {\varepsilon\over
\omega_0'}\sin{\omega_0'\Delta t}$.  Inserting these into the threshold
condition Eq. (\ref{eq:threshold}), we find
\begin{equation}
\varepsilon_c({\Delta t}) = 2 \omega_0' \left( {\cos \omega_0' \Delta t \pm
	\cosh\gamma\Delta t \over {\sin\omega_0'\Delta t}} \right)
\end{equation}
This threshold condition is plotted in Fig. \ref{fig:delta_tongues}a for
different choices of the damping $\gamma$.  The curves generated by choosing
the plus sign are centered on $\Delta t = (n/2) T, n = 1, 3, 5, ...$ and 
are known as subharmonic ``tongues," while the curves generated by choosing
the minus sign are centered on $\Delta t = mT$, $m = 1,2,3, ...$ and 
are known as harmonic tongues.  These curves are qualitatively similar to 
those of the Mathieu Eq. with sinusoidal forcing \cite{benjamin54,kumar94}.

\subsection{Mathieu equation driven by a sequence of delta functions of
alternating sign}
\label{sec:alternatingsigns}

	In our version of the Faraday experiment, the container's vertical
position $x(t)$ is a triangle wave.  Taking
two time derivatives of $x(t)$, we see that the acceleration 
$a(t)$ is a periodic sequence of delta functions of
equal magnitude but alternating sign.  For this reason, we generalize the
analysis of section \ref{sec:equalsigns} to a driving term of the form
\begin{equation}
\varepsilon \left\{ \sum_n \left[ \delta (t-n \Delta t) - 
\delta [t-(n+1/2) \Delta t]\right] \right\}x(t)
\label{eq:doubledelta}
\end{equation}
Let $x_n (t)$ be the solution valid over $(n\Delta t) < t <
[(n+1/2)\Delta t]$ and let $x_{n+1/2}(t)$ be the solution valid over
$[(n+1/2)\Delta t] < t < [(n+1)\Delta t]$.  Except for the kicks at
$(n\Delta t)$ and $[(n+1/2)\Delta t]$, the oscillator is free and 
we have
\begin{mathletters}
\label{eq:m1m2} 
\begin{eqnarray}
x_n(t) &=& A_n e^{\psi (t-t_n)} + {\rm c.c.} \\
x_{n+1/2}(t) &=& A_{n+1/2} e^{\psi (t-t_{n+1/2})} + {\rm c.c.} 
\end{eqnarray}
\end{mathletters}
Applying the continuity and jump conditions at 
$(n\Delta t)$ and $[(n+1/2)\Delta t]$, we find
\begin{mathletters}
\begin{eqnarray}
\left( \begin{array}{c}
A_{n+1}^r \\
A_{n+1}^i
\end{array}
\right) &=& e^{-\gamma \Delta t/2} \left(
\begin{array}{cc}
c & - s \\
s + \alpha c &  c - \alpha s
\end{array} \right)
\left(
\begin{array}{c}
A_{n+1/2}^r \\
A_{n+1/2}^i
\end{array} \right) \\
\vspace{0.25in}
\left( \begin{array}{c}
A_{n+1/2}^r \\
A_{n+1/2}^i
\end{array}
\right) &=& e^{-\gamma \Delta t/2} \left(
\begin{array}{cc}
c & - s \\
s - \alpha c &  c + \alpha s
\end{array} \right)
\left(
\begin{array}{c}
A_n^r \\
A_n^i
\end{array} \right) ,
\end{eqnarray}
\end{mathletters}
where $c \equiv \cos{\omega_0'\Delta t/2}$ and 
$s \equiv \sin{\omega_0'\Delta t/2}$.  Combining these, we can write
$A_{n+1} = e^{-\gamma \Delta t} M_2 M_1 A_n \equiv e^{-\gamma \Delta t} M
A_n$.  Since det$M =$ (det $M_1$)(det $M_2) = 1$, we have the same threshold
condition, $x = \pm \cosh \gamma\Delta t$, leading to 
\begin{equation}
\varepsilon_c (\Delta t) = 2 \omega_0' \sqrt{{\cos\omega_0'\Delta t \pm
	\cosh\gamma\Delta t} \over 1 - \cos\omega_0'\Delta t}
\label{eq:doubledelta_thresh}
\end{equation}
The resonance tongues are shown in Fig. \ref{fig:delta_tongues}b.  There are 
only subharmonic tongues, a feature that turns out to
be special to the use of our particular driving function Eq.
(\ref{eq:doubledelta}).  

\begin{figure}[tbp]
\epsfxsize = 6cm
\epsffile{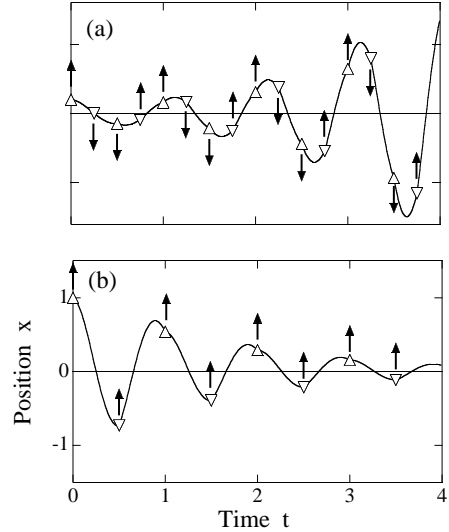}
\caption {Time series $x(t)$ for the Mathieu eq. driven by a periodic sequence
of delta functions of alternating signs.  In both time series, the damping 
$\gamma = 0.1$ and the forcing $\varepsilon = 0.7$.  The triangles show where
the forcing is applied. They point up for positive delta functions and down for negative delta functions.  The arrows above the triangles indicate
the direction of the force vector for each kick.  Because the delta function
forcing term multiplies the coordinate $x$, the sign of the force vector
opposes that of the delta function for $x < 0$.  Time is in units of the 
undamped-oscillator period ($2\pi$). (a) Subharmonic case  
(b) Harmonic case.}
\label{fig:timeseries}
\end{figure}

	To understand more intuitively why there is resonance
for the subharmonic case but not for the harmonic case, consider the two time
series shown in Fig. \ref{fig:timeseries}.  Fig \ref{fig:timeseries}a shows a
time series for $(\Delta t = T/2$, $\omega = 2\omega_0$), where $T = 2\pi/
\omega_0$ is the natural period of the oscillator.  Fig. \ref{fig:timeseries}b
shows the case $(\Delta t = T$, $\omega = \omega_0)$. The delta-function
impulses are indicated by up and down triangles on each curve while the
force vector for each kick is indicated by an arrow above or below the
triangle.  Because the delta functions are multiplied by the coordinate $x$, the sign of the velocity increment depends on the position of the particle relative to the origin.
In the subharmonic case, the impulse always increases the speed of the motion.  In the harmonic case, the up triangles decrease the speed, while the down triangles add to it.  Because the overall motion is damped, it is clear that the net effect of the kicks is to decrease the energy of the system.  One can understand this better by considering the matrices $M_1$ and
$M_2$ in the limit of zero damping (harmonic case).  The cos$\omega_0' \Delta t/2$ and sin$\omega_0' \Delta t/2$ terms are then $-1$ and $0$, respectively, and the product $M_2 M_1$ is just the identity matrix, {\it independent of the value of the forcing} $\alpha$.  Thus, no energy is added to the system even as it is forced arbitrarily hard.  (One kick adds energy while the next one removes it.)  When damping is added, as in Fig. \ref{fig:timeseries}b, the system is always stable about $x = 0$.  The harmonic case differs from the harmonic case of sinusoidal driving, where there is still a resonance because the forcing is distributed over the whole period of the oscillator.

	Note that when $\gamma > 0$, higher order tongues have higher
thresholds than the $\omega = 2\omega_0$ tongue.  The bottom of this first
tongue defines an overall threshold that represents the global-minimum 
amplitude $\varepsilon^*$ required to excite oscillations. 
We can find $\varepsilon^* (\gamma , \omega_0)$ by differentiating 
Eq. (\ref{eq:doubledelta_thresh}) for $\varepsilon(\Delta t)$.  This yields
$\cot \omega_0' \Delta t/2 = {\gamma/\omega_0'} \tanh \gamma\Delta t/2$
for $\Delta t^*(\gamma, \omega_0)$.  This transcendental equation 
can be solved numerically or perturbatively for small
$\gamma/\omega_0$, which gives $\Delta t \approx \pi/\omega_0
[1-(1/2)(\gamma/\omega_0)^2]$ and
$\omega = 2\omega_0 [1+(1/2) (\gamma/\omega_0)^2 ]$.  
To lowest order in $\gamma/\omega_0$, the
resonance conditions are just the zero-viscosity ones: 
$\Delta t = \pi/\omega_0$ and $\omega = 2\omega_0$.  
At this order, the overall forcing threshold is
given by $\varepsilon^* = \pi\gamma$.

\section{Faraday Waves with Triangle-Wave Forcing}
\label{sec:faraday}

	In this section, we shall apply the above results to the
linear-stability analysis of the flat surface of a viscous fluid held in 
a laterally infinite container of fluid of infinite depth.  (The 
finite-depth case is straightforward, but the algebra is longer
\cite{bechhoefer95}.)  Already in 1883, Rayleigh had
suggested that the instability studied fifty years earlier by Faraday might be
due to the phenomenon of parametric resonance, and he wrote down and explored
the properties of Eq. (\ref{eq:mathieu}) independently of Mathieu
\cite{rayleigh1883}.  Seventy years later, Benjamin and Ursell 
\cite{benjamin54} showed that the
normal modes of the linearized, inviscid Navier Stokes equations each obeyed a
Mathieu equation.  In the Benjamin-Ursell calculation, the effects of damping
must be added perturbatively (using $\gamma = 2\nu k^2 /\omega \ll 1$ as the
bulk dissipation coefficient, where $\nu = \eta /\rho$ is the kinematic 
viscosity and $k$ the
magnitude of the wavenumber of the disturbance).  Recently (1994) 
Kumar and Tuckerman performed a complete linear analysis that is valid 
for all viscosities \cite{kumar94}.  Since the analysis for arbitrary 
viscosity is only slightly more complicated than that for small viscosity, we
present the more general result.

	We begin by deriving the equations of motion, following Kumar and
Tuckerman. The Navier-Stokes equations of motion for the fluid layer are
\begin{mathletters}
\begin{eqnarray}
[ \partial_t + \vec{U} \cdot \vec{\nabla}] \vec{U} &=& - {1\over\rho}
        \vec{\nabla} P + \nu \nabla^2 \vec{U} - g(t) \hat{z} \\
\vec{\nabla} \cdot \vec{U} &=& 0 .
\end{eqnarray}
\end{mathletters}
Here, $\vec{U}(x,y,z,t)$ is the fluid velocity field in the lab frame of
reference and $P(x,y,z,t)$ is the pressure field. 
There is a simple solution, corresponding to a
fluid at rest in its own frame of reference:  $\vec{U} = 0$ and
$P(t) = -\rho g(t) z$.  
 
        Linearizing about the at-rest solution, we find for the
perturbing fields $\vec{u}$ and~$p$
\begin{mathletters}
\begin{eqnarray}
\partial_t \vec{u} &=& 
- {1\over\rho} \vec{\nabla} p + \nu \nabla^2 \vec{u} \\
\vec{\nabla} \cdot \vec{u} &=& 0 .
\end{eqnarray}
\end{mathletters}
 (In the new frame of reference, gravity does not appear in the bulk
equations of motion.  It will resurface in the boundary conditions.)
Applying the operator 
(\mbox{$\hat{z} \cdot \vec{\nabla} \times \vec{\nabla} \times$)} 
and using the incompressibility condition $\vec{\nabla} \cdot \vec{u} = 0$,
we eliminate the pressure and obtain
\begin{equation}
(\partial_t - \nu \nabla^2) \nabla^2 u = 0,
\end{equation}
where $u \equiv u_z(x,y,z,t)$ is the $z$-component of the velocity field.

        At the bottom of the cell, no-slip boundary conditions imply that
$\vec{u} = 0$.  Using the incompressibility condition, we rewrite this as
\begin{equation}
u = \partial_z u = 0.
\end{equation}

        At the air-fluid interface, there are three conditions.  The first
is a kinematic one that relates the fluid velocity to the position of the
interface:
\begin{equation}
[\partial_t + \vec{u} \cdot \vec{\nabla}] \zeta =  u |_{z=\zeta},
\end{equation}
where $\zeta(x,y,t)$ is the location of the interface.  To linearize, we
drop
the $\vec{u} \cdot \vec{\nabla}$ term and evaluate the right-hand side
at $z = 0$, giving $\partial_t \zeta = u|_{z=0}$.

        The second condition asserts that there are no tangential stresses
at the interface.  Since the interface normal is just $\hat{z}$ to first
order, one has that $\pi_{xz} = \pi_{yz} = 0$ at $z = 0$, where the 
stress tensor $\pi$ is defined as
\begin{equation}
\pi_{ij}  = -P\,\delta_{ij} + \eta\,(\partial_i u_j + \partial_j u_i )
= \left[ -p + \rho g(t) z \right] \,\delta_{ij} + \eta\,(\partial_i u_j + 
\partial_j u_i),
\label{e:stresstensor}
\end{equation}
with $i$ and $j$ ranging over $x$, $y$, and $z$, and where $\eta$ is
the ordinary fluid viscosity.  Taking the horizontal
divergence of Eq. (\ref{e:stresstensor}), we have
\begin{equation}
\nabla^2_H u + \partial_z (\vec{\nabla}_H \cdot  \vec{u_H}) =
        (\nabla^2_H - \partial_{zz})u = 0,
\end{equation}
where ${\vec{u}_H}$ is the horizontal velocity field at $z=0$ and the
incompressibility condition was used to write the last identity.

        The third condition relates the normal stress to the Laplace
pressure caused by surface tension acting on a curved interface:
\begin{equation}
\pi_{zz} = \gamma \kappa \approx -\gamma \nabla_H^2 \zeta,
\end{equation}
where $\kappa$ is the mean curvature of the interface and the second term
comes from writing the curvature to first-order in $\zeta$.  We have, at $z
= 0$,
\begin{equation}
p(\zeta) = 2 \eta \left( \partial_z u \right) + \rho g(t) \zeta + \gamma
        \nabla_H^2 \zeta.
\label{e:pressure}
\end{equation}
Taking the horizontal divergence of the Navier--Stokes equation, one can
obtain a second relation involving the pressure field and use this to
rewrite Eq. (\ref{e:pressure}) in terms of $u$ and $\zeta$ alone:
\begin{equation}
(\partial_t - \nu \nabla^2) (\partial_z u) = 2\nu \nabla_H^2
(\partial_z u) + g(t) \nabla_H^2 \zeta - {\gamma\over\rho} \nabla_H^4 \zeta.
\end{equation}

        Because the horizontal eigenfunctions of a laterally infinite
container are $e^{\pm i\vec{k}\cdot\vec{r}}$, we can replace
$\nabla_H^2$ by $-k^2$, and $\nabla^2$ by $(\partial_{zz} - k^2)$.
We then have the following linearized equations of motion for $u$:

\end{multicols}
\widetext
\vspace{3mm}\hrule\vspace{3mm}

\vbox{
\begin{mathletters}
\label{eq:eqsofmotion}
\begin{eqnarray}
\left[ \partial_t - \nu (\partial_{zz} - k^2) \right] 
	( \partial_{zz} - k^2) u &=& 0 \\
u &=& \partial_z u = 0    \hspace{1in} z = -\infty \\
( \partial_{zz} + k^2) u \left|_{z=0} \right. &=& 0  \hspace{1.5in} z = 0 \\ 
\label{eq:eqsofmotiond}
\left[ \partial_t - \nu (\partial_{zz} - k^2) + 2\nu k^2 \right] 
	(\partial_z u)|_{z=0}
&=& - \left[ g[1 - f(t)] - {\gamma \over \rho} k^3 \right] k^2 \zeta \\ 
\partial_t \zeta &=& u |_{z=0}
\end{eqnarray}
\end{mathletters}}
As with the Mathieu eq., one can easily solve the unforced equations between
kicks, where surface waves propagate and are damped by viscosity.  Let
$u_n(z,t)$ be the $z$-component of the velocity valid for $(n\Delta t) < t <
(n+(1/2)\Delta t)$.  Let $\zeta_n(t)$ be the amplitude of surface displacement
over the same time period.  From Eqs. (\ref{eq:eqsofmotion}), we find
\begin{mathletters}
\label{eq:un_zetan}
\begin{eqnarray}
u_n(z,t) &=& e^{\psi (t - t_n)} u_n(z) + {\rm c.c.} \hspace{1in}
u_n(z) = a_n e^{\kappa z} + b_n e^{kz} \\
\zeta_n(t) &=& \zeta_n e^{\psi (t-t_n)} + {\rm c.c.} \hspace{1.25in} 
\kappa^2 = k^2 + \psi/\nu
\end{eqnarray}
\end{mathletters}

\vspace{6mm}
\begin{multicols}{2}

Eqs. (\ref{eq:un_zetan}) give $a_n = -2\nu k^2 \zeta_n$ and 
$b_n = (2\nu k^2 + \psi ) \zeta_n$.  
The normal-stress condition (Eq. \ref{eq:eqsofmotiond}) gives
the dispersion relation $\psi (k)$ for the surface waves: 
\begin{equation}
\psi = - 2\nu k^2 \pm i\omega_0\sqrt{1 - {4\nu^2k^3\kappa\over {\omega_0^2}}}, 
\label{eq:dispersion}
\end{equation}
where $\omega_0^2 \equiv gk + (\gamma /\rho ) k^3$.
Note that this expression for $\psi$ resembles the one found for the Mathieu
equation.  (See Eq. \ref{eq:solution} and the line following.)
The laterally infinite container supports a continuum of modes 
indexed by the 2D wavenumber $\vec{k} = (k_x,k_y)$.  
A mode of wavenumber $k$ has a damping coefficient $\gamma \approx 2\nu k^2$. 
Damping gives a (second-order) decrease in the frequency, so that 
$\omega_0' = \omega_0 \sqrt{1 - {4\nu^2 k^3 \kappa \over \omega_0^2}}.$  
One complication, relevant
only for higher viscosities, is that Eq. (\ref{eq:dispersion}) is an implicit
equation for $\psi$, since $\kappa = \kappa (\psi )$.  One can write (and
solve) an explicit, fourth-order equation for $\psi (k)$, but not much insight
is gained.  The important point is that $\psi (k)$ is the dispersion relation
for free surface waves propagating on a deep, viscous fluid.  Such dispersion
relations have been catalogued for a variety of physical situations 
(small or large depth,
small or large viscosity, viscoelastic effects, surfactant layers on the
surface, etc.).  One can look up or calculate the appropriate $\psi (k)$ for
each physical situation one wishes to study.  The dispersion relation is 
completely independent of the way the waves are excited (parametric or 
direct forcing).  For now, we leave $\psi (k)$ as arbitrary.

	The next step is to derive the matrices relating the complex
amplitudes $\zeta_{n+1}$ to $\zeta_{n+1/2}$ to $\zeta_n$. 
At $(n+1/2)\Delta t$, continuity of $\zeta$ yields Re$(\zeta_{n+1/2} ) = $ 
Re$(\zeta_n e^{\psi \Delta t/2})$.  The velocity-jump condition is derived by 
integrating the normal-stress condition Eq. (\ref{eq:eqsofmotiond}) over 
a small interval of time centered on $(n+1/2)\Delta t$.  We find

\end{multicols}
\widetext
\vspace{3mm}\hrule\vspace{3mm}

\begin{equation}
(\partial_z u_{n+1/2}|_{z=0}) (t_{n+1/2}) - (\partial_z u_n|_{z=0}) (t_{n+1/2})
 = {4x_{pp}k^2 \over {\Delta t}} (-1) \left[ \zeta_n e^{\psi\Delta t/2} +
{\rm c.c.} \right]
\label{eq:fluidjump}
\end{equation}
where
\begin{equation}
\left( \partial_z u_{n+1/2}  \right) (t_{n+1/2}) =
\kappa a_{n+1/2} + k b_{n+1/2} + \rm{c.c.} = 
[2\nu k^2(k-\kappa ) + \psi k] \zeta_{n+1/2} + \rm{c.c.}
\end{equation}
Inserting this into Eq. (\ref{eq:fluidjump}), we find
\begin{equation}
{\rm Im} \zeta_{n+1/2} = {\rm Im} \zeta_n e^{\psi \Delta t/2} + {4x_{pp} k
\over {\Delta t}} {1\over{({\rm Im} \psi - 2\nu k {\rm Im}\kappa})}
{\rm Re}\zeta_n e^{\psi \Delta t/2}
\end{equation}

\vspace{3mm}\hrule\vspace{3mm}
\begin{multicols}{2}

Define now $\psi \equiv -\gamma + i \omega_0'$ and $\kappa = \kappa_r +
i\kappa_i$.  Then
$\zeta_{n+1} = e^{-\gamma\Delta t/2} M_2 \zeta_{n+1/2}$ and 
$\zeta_{n+1/2} = e^{-\gamma\Delta t/2} M_1 \zeta_n$. 
Here, the matrices $M_1$ and $M_2$ have exactly the same form as before,
except that now $\alpha = 4 x_{pp}k/[(\Delta t)(\omega_0' - 2\nu k
\kappa_i)]$.  The tongues then are given by
$2\alpha({\Delta t}) =  \sqrt{{\cos\omega_0'\Delta t \pm
\cosh\gamma\Delta t} \over 1 - \cos\omega_0'\Delta t}$.
One difference is that in a laterally infinite container, all modes $\vec{k}$
are available.  Thus, whatever the driving frequency $\omega$ (or whatever
the $\Delta t$), the response of all modes must be considered.  
This implies that the overall threshold $\varepsilon^*$ discussed 
above is the relevant concept. Since the $\omega = 2\omega_0$ tongue 
is in general the easiest to excite, one expects waves at just half the 
driving frequency. 

	For small damping $(\sigma \equiv 4\nu k^2 /\omega \ll 1)$, 
one has $\gamma \approx 2\nu k^2$, $\kappa = k$, 
$\omega_0' \approx \omega \approx 2\omega_0$, giving a threshold of
\begin{equation}
a_{pp} = \left( {\pi^2\over 2} \right) \nu k \omega ,
\end{equation}
which is slightly higher than the corresponding condition 
for sine-wave forcing \cite{fauve92}, $a_{pp} = 4\nu k^2 \omega.$

\section{Driving by an Asymmetric Triangle Wave}
\label{sec:2freq}

	Recently, Edwards and Fauve \cite{edwards94} have explored the
effects of exciting Faraday waves with a driving signal that is the sum of two
sine waves of rationally related frequency.  They showed that for fixed
frequency ratio (5:4 in their original work), the relative amplitude and phase
of the two sine waves could be considered as independent control parameters. 
Varying these new parameters dramatically affects the pattern-formation
selection.  For example, they produced hexagons and quasi-periodic patterns
(``quasi-patterns") under conditions where single-frequency forcing would lead
to a stripe pattern.

	Here, we look at a simple model of two-frequency forcing, an
asymmetric triangle-wave forcing (Fig. \ref{fig:driving}c).  This
waveform has the crucial feature that it is asymmetric under time reversal, 
$f(t) \neq f(\Delta t - t)$, 
in contrast with the ``single-frequency" driving waveform
(Fig. \ref{fig:driving}b).  As Edwards and Fauve have argued, time-reversal
invariance of $f(t)$ implies that the nonlinear amplitude equations for the
parametrically excited waves must be invariant under $A \to -A$; however, this
symmetry is removed for asymmetric $f(t)$.

	In the delta-function model, asymmetric triangle-wave forcing is
easily solved by replacing the $\Delta t/2$ in the matrices $M_1$
and $M_2$ (Cf. Eqs. \ref{eq:m1m2}) with $\Delta t_1$ and $\Delta t_2$,
where $\Delta t_1 + \Delta t_2 = \Delta t$. 
The tongue boundaries for the Mathieu Eq. are then given by
\begin{equation}
\varepsilon_c({\Delta t}) = \sqrt{2} \omega_0' \sqrt{{\cos\omega_0'\Delta t \pm
\cosh\gamma\Delta t} \over 
{\sin\omega_0'\Delta t_1  \sin\omega_0'\Delta t_2}} .
\end{equation}
The tongues are plotted for several choices of $\Delta t_1$ and $\Delta t_2$ in
Fig. \ref{fig:delta_tongues}c.  
Note the reappearance of the harmonic tongues for $\Delta t_1 \neq \Delta t_2$.  In this simple model, there is no analog of the
``bicritical point" discussed by Edwards and Fauve, where two different
wavenumbers go unstable simultaneously.  Nonetheless, the symmetry argument
given above suggests that asymmetric-triangle-wave forcing will alter the
nonlinear pattern selection in much the same way as two frequencies do.

\section{Conclusion}
\label{sec:conclusion}

	In this paper, we have shown how the linear-stability analysis of 
Faraday waves is greatly simplified for triangle-wave forcing.
Our analysis is simple enough to include in an undergraduate hydrodynamics
course.  One could also combine the analysis with a simple experiment to
make a good senior thesis project.  References \cite{edwards94} and
\cite{bechhoefer95} give careful discussions of the required experimental
apparatus \cite{exp_note}.  

	Our calculation may also serve as the starting point for a more extensive
nonlinear analysis.  To date, nonlinear amplitude equations for Faraday waves
have been derived only in the limit of small damping 
$\sigma \ll 1$ \cite{milner91,zhang94};
however, many interesting phenomena are seen when $\sigma \approx 0.4$ to
$0.7$ \cite{daudet}.  
With delta-function driving, the effects of the parametric pumping are
straightforward, even in the nonlinear regime.  The hard part of the
calculation becomes the determination of nonlinear terms in the dispersion
relation for surface waves.  However, this is a problem that has received much 
attention over the years, and a number of results are available 
\cite{debnath94}.

\section*{Acknowledgments}

This work was supported by NSERC (Canada).

\end{multicols}						         

\begin{references}
\bibitem{exp_note} The electromagnetic shaker tables that are usually used 
to generate Faraday waves will not give an accurate output if a 
function generator is used to create the triangle wave.  
Because of the inductance of the vibration coils, the different frequency 
components of the input waveform will have different phase shifts and the
result will be far from a triangle wave.  A straightforward work-around is to 
use the feedback system of \protect{\cite{edwards94}}, which requires a 
modest amount of programming.  Alternatively, one could design a 
vibrator specifically for triangle waveforms.
\end{references}
\end{document}